# Noble gas functional defect with unusual relaxation pattern in solids

Lovelesh[1*], Harshan Reddy Gopidi[1*], Muhammad Rizwan Khan[1], and Oleksandr I. Malyi[1,#]

[1]Centre of Excellence ENSEMBLE[3] Sp. z o. o., Wolczynska Str. 133, 01-919, Warsaw, Poland

*the authors contributed equally to this work
#Email: oleksandrmalyi@gmail.com

**Abstract:**
The conventional understanding has always been that noble gases are chemically inert and do not affect materials properties. This belief has led to their use as a standard reference in various experimental applications through noble gas implantation. However, in our research, using first-principles calculations, we delve into the effects of noble gas defects on the properties of several functional oxides, thereby questioning this long-held assumption. We provide evidence that noble gases can indeed serve as functional defects. They have the potential to decentralize the localized defect states and prompt a shift of electrons from a localized state to the main conduction band. Our investigation unveils that noble gas defects can indeed significantly alter material properties. Thus, we underscore the importance of factoring in such defects when assessing material properties.

***Electronic structure theory as a common tool to investigate defect physics in solids:*** Point defects, which originate from crystal growth, ion implantation, change of environmental conditions, or simply intrinsic increase of configurational entropy, can significantly affect materials properties. This is notably pronounced in wide band gap insulators, where defects play a major role in determining the equilibrium Fermi level, doping behavior, and material color.[1-3] With the development of electronic structure theory, it has become clear that properties of point defects (e.g., formation energies, concentration, and transition levels) can be well understood from density functional theory (DFT). However, since DFT calculations are constrained in supercell size (conventional setup involves typically only a few hundred atoms), the results can be directly affected by the periodic boundary conditions employed in DFT software (e.g., Vasp[4-8], Quantum Expresso[9], and Siesta[10]). In practice, it is assumed that defect-induced local deformation can be well minimized in supercells having effective lateral dimensions larger than 10 Å in each direction while the other corrections (e.g., charge-charge interactions in supercells containing charged defects[11-13] and band filling[14]) can be performed as part of post-processing procedures. Such beliefs are supported by the development of different post-process codes like PyLada-defects[15], AiiDA-defects[16], and PyCDT[17] to help mitigate these challenges.

***Noble gas as a functional defect in solids:*** Transitioning to a specific yet intriguing category of defects, noble gases, we witness a remarkable break from convention. The prevailing assumption is that noble gases do not impact the properties of materials. Consequently, in procedures such as sputtering or wet-chemical synthesis within a glove box, the presence and potential impact of noble gases are typically disregarded. Additionally, noble gas implantation (with Ar)[18-21] is usually used as a reference state to eliminate the effects of implantation damage. This technique specifically relies on the assumption that Ar defects do not modify the material properties. Although noble gases are considered chemically inert due to their fully occupied orbitals, they can still influence material properties through repulsion with atomic orbitals of surrounding atoms. This repulsion can result in changes to the charge density distribution, effectively making the noble gas a functional defect.[22] Such unique behavior thus breaks the common belief that chemically inert defects cannot be functional defects in solids. However, a fundamental question remains: does the noble gas have a substantially different effect on other material properties beyond functionality? Motivated by this, in this work, we revisit the basic physics of the formation of noble gas defects in solids using the example of $SnO_2$, ZnO, and $ZrO_2$. Through a series of first-principles calculations, we demonstrate that introducing noble gases into the material lattice structure can result in the delocalization of localized stat es and unusual relaxation patterns not typical for common defects. Specifically, when noble gas occupies the vacancy site, it can delocalize the localized defects states (if one is located on the vacancy site), i.e., moving electrons from a localized in-gap state to the principal conduction band, this is fully caused by the repulsion of atomic orbitals. This work highlights the unusual role that noble gases can play in solid-state physics, raising new questions about the nature of noble gas defects and their impact on material properties.

**Physics of localized in-gap defect states in solids:** The formation of point defects in functional materials often leads to the creation of localized in-gap states, which are electronic states located within the energy range between the valence and conduction bands. These localized in-gap states typically emerge due to the ability of electrons or holes to become localized at lattice imperfections, such as a vacancy site, or through the formation of a polaronic state localized on one of the sublattices.

In the past, these states have not been heavily distinguished in electronic structure research; however, they carry significant consequences for material functionality. For instance, recently, their role in antidoping behavior has been demonstrated[23-27], highlighting their potential impact on the electrical and optical properties of materials. Furthermore, the presence of localized in-gap states can influence electron/hole transport[28-29], which is a critical parameter for the performance of electronic and optoelectronic devices. In some cases, these states can also serve as recombination centers, limiting the efficiency of solar cells, light-emitting diodes, and other semiconductor-based devices.[30-32] Recent advances in experimental techniques (e.g., different spectroscopic techniques) have allowed for a more detailed investigation of these localized states, shedding light on their origin, nature, and potential applications.[33-38] Fig. 1a,b illustrates a scenario where a donor defect triggers the development of an occupied in-gap state near the valence band, as exemplified by $SnO_2:V_O$ (oxygen vacancy in $SnO_2$). The analysis of partial charge density indicates that the defect states are occupied by 2e and are localized primarily on the vacancy site, with a minor contribution from the surrounding atoms. The same nature of oxygen defects is also known for other functional oxides (e.g., ZnO[22, 39] and $ZrO_2$[40]).

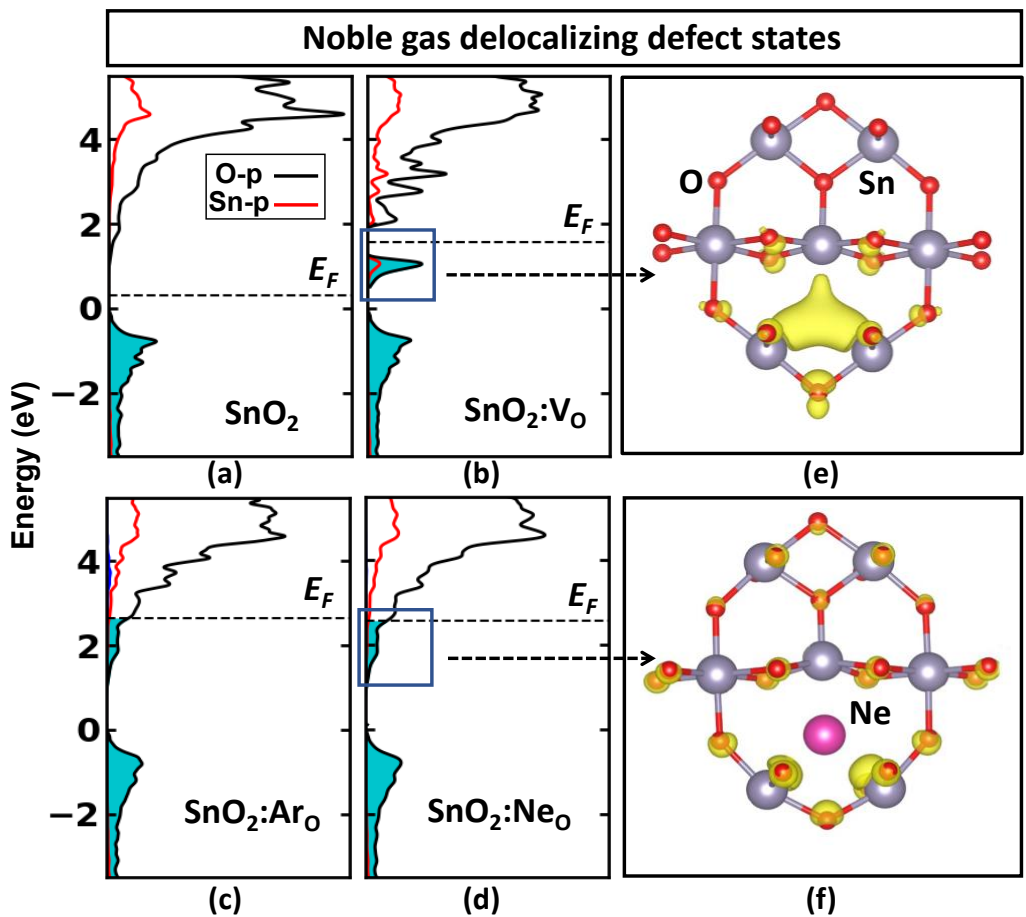

**Figure 1: Noble gas defect resulting in the delocalization of in-gap defect states localized on vacancy site.** Electronic density of states for (a) $SnO_2$, (b) $SnO_2:V_O$, (c) $SnO_2:Ar_O$, and (d) $SnO_2:Ne_O$. For visualization purposes, the density of states above bulk like region are multiplied by 50. The occupied states are shown as shadow regions. (e) Partial charge density corresponding to the localized state for $SnO_2:V_O$. The isosurface is set at 7 m$e$/Å³. (f) Partial charge density corresponding to the occupied part of the conduction band for $SnO_2:Ne_O$. The isosurface is set at 3 m$e$/Å³. All results are presented for 72-atom $SnO_2$ supercell. In the insulating state, the Fermi level is place midway between the highest occupied state and the lowest unoccupied state

**Unexpected physics of noble gas interaction with localized in-gap states:** To understand better the physics/chemistry of noble gas defects, we investigate the interactions between Ne/Ar defects with $SnO_2:V_O$ and $ZrO_2:V_O$ systems. In both systems, noble gases can be located either far from the defect (i.e., oxygen vacancy) or at the vacancy site. The precise behavior largely hinges on the reduction of strain induced by the noble gas defect and its consequent influence on electronic properties. While the primary focus of this research does not lie on an exact comparison of defect formation energy as a function of the distance between the noble gas and the vacancy, it is important to note that in many scenarios, noble gas defects opt for the vacancy site, which presents itself as the lowest energy configuration.[22] When noble gas defects locate away from the vacancy site, their impact on electronic properties is relatively insignificant, unless they disrupt (e.g., via breaking bonds) the chemical bond between the host atoms. Conversely, when a noble gas occupies the vacancy site, there is a shift of electrons from the occupied in-gap state to the principal conduction band - essentially, two electrons are transitioned to the conduction band while the localized in-gap states vanish. This observation holds true for $SnO_2:Ne_O$, $SnO_2:Ar_O$, and $ZrO_2:Ne_O$ cases and, indeed, does not look to be sensitive to the noble gas element (as illustrated in Fig. 1 on an example of $SnO_2:Ne_O$ and $SnO_2:Ar_O$ systems). At a cursory glance, one might presume that a noble gas defect contributes electrons to surrounding atoms, thereby functioning as a typical defect. However, this behavior is rather atypical and physically uncharacteristic for noble gases as well as cannot explain the disappearance of the occupied in-gap states. Moreover, as the existing literature[41-42] indicates, noble gases are primarily known to exhibit polarization when approaching material surfaces and are not recognized as chemically active species.

To explore the underlying physics/chemistry of this interaction, we examine the charge density redistribution caused by the insertion of a noble gas at the vacancy site. This is all done by calculation of charge density difference (i.e., $\Delta\rho = \rho(host:NG) - \rho(host) - \rho(NG)$, where $\rho(host:NG)$, $\rho(host)$, and $\rho(NG)$ are charge density for the host system containing the noble gas defect, host system (e.g., $SnO_2:V_O$), and noble gas all calculated for atomic positions in host system containing noble gas) and its spatial dependence. Here, the spatial dependence analysis is done by analyzing the local $\Delta\rho(r)$ distribution around the noble gas atom as the function of the radius. Figure 2 displays the results for $SnO_2:Ne_O$ and $ZrO_2:Ne_O$ systems, which show a slight increase in the charge density around the noble gas atom due to its positioning at the vacancy site. This enhancement occurs within extremely close proximity to the noble gas atom, potentially giving the impression of an evident charge transfer to the noble gas atom. However, a closer inspection reveals that the charge density in the vicinity of neighboring host atoms also elevates, while the charge density between Ne and host atoms significantly decreases. This behavior contrasts with the norms of bond chemistry.[43-44] For instance, in typical scenarios, the formation of a covalent bond would be expected to amplify charge density between atoms, while the formation of an ionic bond would result in transfer of an electron from a less electronegative atom to a more electronegative atom. The calculated results can be understood as follows: noble gases neither accept nor donate electrons, they simply induce repulsion between their atomic orbitals and surrounding atoms. Yet, due to this orbital repulsion, the noble gas can induce delocalization of the localized state in its close vicinity, as the noble gas tends to reduce the overlap of its charge density with surrounding atoms. Remarkably, this scenario proves compatible across all aforementioned systems indicated systems above (i.e., $SnO_2:V_O$, $ZrO_2:V_O$, and $ZnO:V_O$) where the in-gap states are localized on the vacancy site. This understanding further introduces the potential for localized state physics, unaffected by noble gases. As such, if the defect state localization does not coincide with the insertion site of noble gas, the presence of a noble gas defect will not trigger a state delocalization. This scenario can, quite simply, manifest in instances where there is a substantial separation between the noble gas site and the localization of charge density. Such instances include a large spatial separation between the vacancy and the noble gas or when the localized state exists on the atomic sublattice. Our findings definitively showcase the role of noble gas defects as functional defects. They can indeed initiate significant modifications in material properties. However, it is worth noting that under standard conditions, the existence of a high concentration of noble gas defects is unlikely. This improbability mainly stems from the fact that the formation of point defects does not facilitate the necessary chemical bonding required to reduce defect formation energy. Yet, these observations underline a crucial point - when a material sample is exposed to noble gas bombardment, such as through ion implantation, the resultant noble gas defects can drastically change material properties. In such situations, experimentalists and theoreticians should be aware of the possible functionality of noble gas defects.

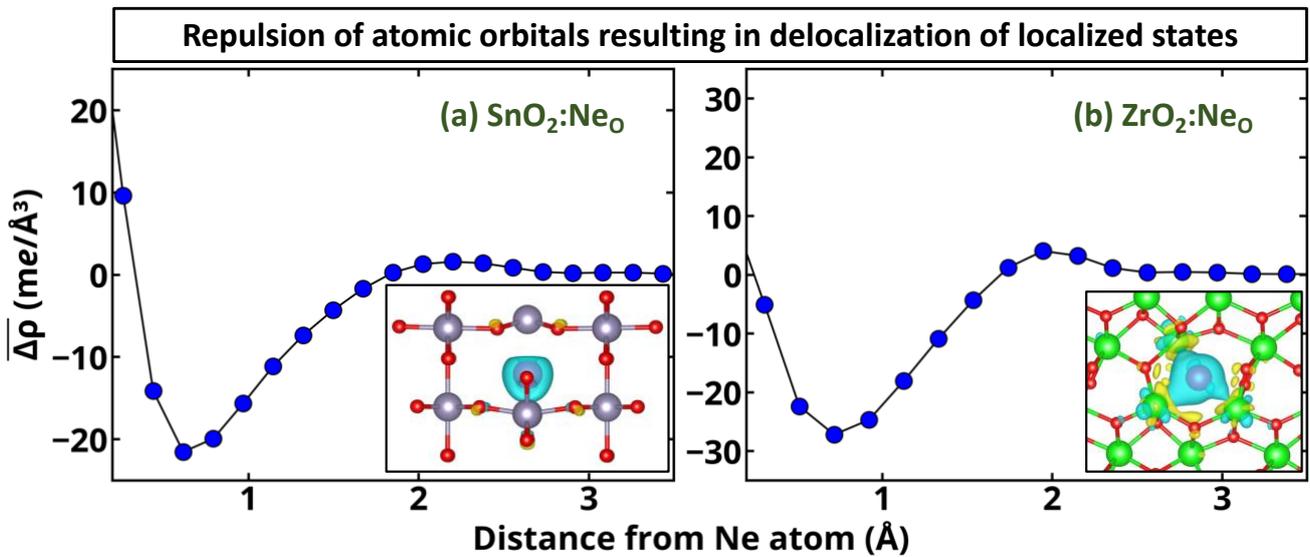

**Figure 2. Repulsion of atomic orbitals as a mechanism to delocalize localized defect state:**
Radial distribution of charge density difference for (a) $SnO_2:Ne_o$ and (b) $ZrO_2:Ne_o$. The corresponding charge density difference (i.e., $\Delta \rho = \rho(host:NG) - \rho(host) - \rho(NG)$, where $\rho(host:NG)$, $\rho(host)$, and $\rho(NG)$ are charge density for the host system containing the noble gas defect, host system (i.e., $SnO_2:V_o$ or $ZrO_2:V_o$), and noble gas all calculated for atomic positions in host system containing noble gas) is shown as insets. The isosurface is shown at 2 m$e$/Å$^3$. The yellow and blue regions correspond to charge density increase and charge density reduction, respectively. The calculations for $SnO_2$ and $ZrO_2$ are performed 72- for 324-atom supercells, respectively.

**Unusual relaxation pattern in solids:** Traditionally, it is believed that the formation of the point defect only requires minimization of charge-charge interaction (present in charge defect calculations), potential alignment correction, band-filling correction, and band gap correction[14]. This practically means that extending the supercell (in all directions) behind 10 Å and simple application of post-process correction is usually sufficient to describe defect properties with high accuracy. Such statements directly assume that other types of interactions are minimized. We note, however, that there are some substantial differences in the noble gas defect case, as the formation of point defect mainly results in strong repulsion of atomic orbitals leading to significant strain. To illustrate this behavior, we consider the case of the $ZnO:He_O$ defect, where He occupies the vacancy site, in comparison with other donor defects in $ZnO:Ga_{Zn}$ and $ZnO:F_O$. All these defects act as donors resulting in putting the Fermi level in the conduction band (unless defect compensation is considered). What makes the situation substantially different between them is the relaxation patterns. To demonstrate this, we analyze the distribution of local atomic displacements and their relative energetics (Fig. 3a,b). Surprisingly, even for relatively large supercell sizes (i.e., one having effective lattice constants larger than 10 Å), there is a relatively large energy lowering with the increase of supercell size. For instance, we find that for ZnO:He$_O$, the relative relaxation energy (i.e., the relaxation energy relative to the largest supercell) lowers by about 0.62 eV as supercell size increases from 96 atoms to 640 atoms. This energy is significantly larger as compared to that for ZnO:F$_O$ and ZnO:Ga$_{Zn}$, where the corresponding value is only about 0.1 eV. The difference is directly related to structural changes. Thus, as expected, the formation of the He$_O$ defect in ZnO results in pushing out surrounding Zn atoms (i.e., atoms within the first coordination sphere, shown schematically as an inset in Fig. 3b). Thus, for 96 atom supercell, the nearest Zn atoms are at a distance of 2.55 Å, while for 640-atom supercell the corresponding distance is 2.59 Å. This results in the change of relative atomic displacement of around 0.04 Å, which is significantly higher than that in ZnO:F$_O$ and ZnO:Ga$_{Zn}$ cases (Fig. 3b). These findings corroborate the idea that noble gases, rather than merely acting as functional defects, can also induce unique relaxation patterns dissimilar to those triggered by traditional defects. We attribute these distinctive characteristics solely to the nature of the interactions instigated by the defect in the system. Contrary to other donor defects, the introduction of a noble gas defect provokes a repulsion of atomic orbitals—a phenomenon that necessitates significantly larger supercell dimensions for accurately calculating both the defect formation energy and the surrounding atomic displacements. Taking into account that He is the smallest considered atom, it is expected that for other noble gas defects (i.e., larger ones), this relative energy and corresponding atomic displacements are even larger.

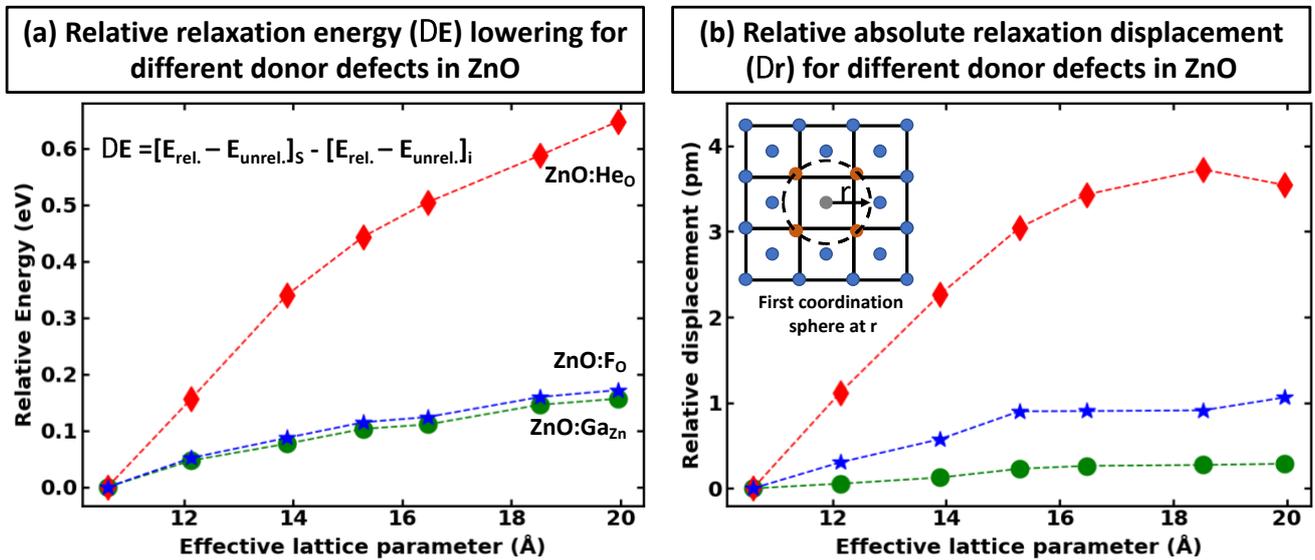

**Figure 3: Supercell size-dependent properties of donor defects in ZnO:**
(a) Relative relaxation energy (i.e., computed as difference in relaxation energy for given supercell with respect to the smallest cell used in this work) for ZnO:Ga$_{Zn}$, ZnO:F$_O$, and ZnO:He$_O$ defects as a function of supercell size. (b) Relative absolute value of relaxation displacement for first coordination sphere atoms around the defect as a function of supercell size. The effective lattice parameter is calculated as the cubic root of the volume.

In summary, this study challenges traditional beliefs about the role of noble gases in solid materials. Despite their chemical inertness, noble gases have been shown to function as defects in solids, causing changes in the charge density distribution, and, subsequently, significantly altering the material properties. This discovery subverts the prevailing notion that chemically inert defects cannot function as such in solids as noble gases can cause the delocalization of localized defect states. Specifically, they can shift electrons from a localized in-gap state to the principal conduction band. This behavior is expected for the systems where the localized state is located on the vacancy site for instance. Interestingly, noble gases neither accept nor donate electrons; rather, they induce repulsion between their atomic orbitals and surrounding atoms, leading to the delocalization of the localized state. We also show such unique interaction can induce unusual relaxation patterns where the relaxation is significantly strong than for typical defects (often requiring convergence with lateral sizes larger than 20 Å). While the formation of a large concentration of noble gas defects is unlikely under normal conditions, when a material sample is bombarded by noble gas, such as during ion implantation, the effects of noble gas defects on material properties cannot be overlooked. This study, thus, underscores that noble gas defects cannot be ignored and indeed be functional defects.

**Methods:** All spin polarized first-principles calculations were carried out using the Vienna Ab initio Simulation Package (VASP)[4-8] and the Perdew-Burke-Ernzerhof (PBE)[45] functional. The cutoff energies for plane wave basis are set to 550 and 450 eV for volume relaxation and final static calculations, respectively. For $SnO_2$, ZnO, and $ZrO_2$ the calculations are performed for tetragonal, hexagonal and monoclinic crystal structures. All supercells were relaxed until the internal forces were smaller than 0.01 eV/Å. The results are presented for Γ-centered Monkhorst–Pack k-grids[46] scheme with k-grid density of 10,000 per reciprocal atom. The results were analyzed using pymatgen[47] and Vesta[48].

**Acknowledgment:** The authors thank the "ENSEMBLE[3] - Centre of Excellence for nanophotonics, advanced materials and novel crystal growth-based technologies" project (GA No. MAB/2020/14) carried out within the International Research Agendas programme of the Foundation for Polish Science co-financed by the European Union under the European Regional Development Fund and the European Union's Horizon 2020 research and innovation programme Teaming for Excellence (GA. No. 857543) for support of this work. We gratefully acknowledge Poland's high-performance computing infrastructure PLGrid (HPC Centers: ACK Cyfronet AGH) for providing computer facilities and support within computational grant no. PLG/2023/016228.